\documentclass[prb,superscriptaddress,twocolumn,floatfix]{revtex4}
\usepackage{graphicx}

\begin{document}

\title{Interplay of structural and electronic phase separation in
single crystalline La$_{2}$CuO$_{4.05}$ studied by neutron and
Raman scattering}

\author{V. P. Gnezdilov}
\affiliation{B. I. Verkin Institute for Low Temperature Physics NASU,
61164 Kharkov, Ukraine}
\author{Yu. G. Pashkevich}
\affiliation{A. A. Galkin Donetsk Phystech NASU, 83114 Donetsk, Ukraine}
\author{J. M. Tranquada}
\affiliation{Physics Department, Brookhaven National Laboratory, Upton, NY
11973, USA}
\author{P. Lemmens}
\affiliation{Max-Planck-Institut f\"{u}r Festk\"{o}rperforschung, 70569
Stuttgart, Germany}
\affiliation{2. Physikalisches Institut, RWTH Aachen, 52056 Aachen,
Germany}
\author{G. G\"{u}ntherodt}
\affiliation{2. Physikalisches Institut, RWTH Aachen, 52056 Aachen,
Germany}
\author{A. V. Yeremenko}
\affiliation{B. I. Verkin Institute for Low Temperature Physics NASU,
61164 Kharkov, Ukraine}
\author{S. N. Barilo}
\author{S. V. Shiryaev}
\author{L. A. Kurnevich}
\affiliation{Institute of Physics of Solids \&\ Semiconductors, Academy of
Sciences, 220072 Minsk, Belarus}
\author{P. M. Gehring}
\affiliation{NIST  Center  for  Neutron  Research,  National  Institute  of
Standards and Technology, Gaithersburg, Maryland 20742, USA}

\date{\today }

\begin{abstract}
We  report  a neutron and Raman scattering study of a
single-crystal of La$_{2}$CuO$_{4.05}$ prepared by high
temperature electrochemical oxidation.  Elastic neutron scattering
measurements  show the presence  of two phases, corresponding  to
the  two edges of the first miscibility gap, all the way up to 300
K. An additional oxygen redistribution, driven by electronic energies,
is identified at 250 K in Raman scattering (RS) experiments by  the
simultaneous onset of two-phonon and  two-magnon scattering, which
are fingerprints of the insulating phase. Elastic neutron scattering
measurements show directly an antiferromagnetic ordering below a N\'eel
temperature of $T_{N} =210$~K. The opening of the superconducting gap
manifests itself as a  redistribution of electronic Raman scattering
below the superconducting transition temperature,  $T_c = 24$~K. A
pronounced temperature-dependent suppression of the intensity  of the
(100) magnetic Bragg peak has been detected below $T_c$. We ascribe this
phenomenon to a  change  of  relative volume fraction of superconducting
and antiferromagnetic phases  with decreasing temperature caused  by  a
form  of a superconducting proximity  effect.
\end{abstract}

\maketitle

\section{INTRODUCTION}

The development of different kinds of charge inhomogeneities at low
temperature is a well known intrinsic property of many
strongly-correlated-electron systems.  Recent studies of Matsuda
\emph{et al.}\cite{Matsuda} on very lightly doped
La$_{2-x}$Sr$_x$CuO$_4$ show that even at $x < 0.02$ the system
demonstrates an electronic phase separation in which regions
of hole-rich and hole-free phases are formed. This observation
indicates the possibility for charges to create nonuniform
phases\cite{Emery} in spite of low doping and an expected uniform
distribution of Sr-dopants.

Oxygen-doped La$_{2}$CuO$_{4+\delta}$ presents a system in which
charge inhomogeneities exist on the background of phase-separated
oxygen interstitials,\cite{Jorg} with an intimate interrelation of
structural (excess oxygen) and charge subsystems. The
discovery\cite{gran87} of high-$T_c$ superconductivity in samples of
nominally pure La$_{2}$CuO$_4$ led to the first evidence of charge
disproportionation in this  material.  The connection between
superconductivity and the hole-rich phase with a finite concentration of
interstitial oxygen was directly proved by annealing in high
pressure oxygen.\cite{schi88} It was reported that after annealing, the
superconducting volume fraction had been increased.\cite{Jorg} Although
the average concentration of charge carriers, roughly estimated as $p
\sim 2\delta$, was moderate, the relatively high temperature of the
superconducting phase transition,
$T_c=32$~K, was achieved even at $\delta=0.05$.

Many investigations of La$_{2}$CuO$_{4+\delta}$ have focused on the
structural aspects of the phase separation phenomena  (for reviews, see
Refs.~\onlinecite{Proc,JMT_rev}). The powder neutron diffraction
measurements by Jorgensen \emph{et al}.\cite{Jorg} showed that at low
temperature, samples contained two very similar orthorhombic
phases,  a primary one with $\delta \approx 0$, and a second,
oxygen-rich phase that was superconducting. At high temperatures,
only a single phase was present, with a reversible
oxygen phase separation occurring (in one sample) at 320~K.
The development of electrochemical-oxidation
techniques\cite{Grenier,Chou1,Radael1} has made it possible to determine
the $T$-$\delta$ phase diagram of La$_{2}$CuO$_{4+\delta}$, the most
prominent feature of which is a miscibility gap in the
region\cite{Radael2} $0.01< \delta < 0.055$. Elastic neutron scattering
from large electrochemically-oxygenated single crystals revealed a
new kind of periodicity in oxygen-rich regions---an ordering of
the interstitially-occupied oxygen layers along the $c$-axis, commonly
referred to as staging behavior.\cite{Wells,Xiong} These results indicated
that the phase diagram of La$_{2}$CuO$_{4+\delta}$ at higher
oxidation levels includes new miscibility gaps separating regions of
staged phases.\cite{Birgeneau}

It is generally presumed that the first miscibility gap is driven by a
tendency for holes doped into antiferromagnetic CuO$_2$ planes to phase
separate.\cite{Emery}  The phase separation is screened by the mobile
oxygen interstitials.  The existence of hole-poor and hole-rich phases at
lower temperatures is demonstrated by the observations of, respectively,
antiferromagnetic order and superconductivity\cite{ansa89,hamm90,vakn94};
however, the ordering temperatures are lower than the structural phase
transition temperature for the interstitials.  One motivation for the
present work is to test more directly the connection between the
electronic and interstitial phase separation transitions through the use
of Raman scattering, which is sensitive to the presence of a
correlated-insulator phase through two-magnon and two-phonon scattering.

Another aspect that we explore is the effect of quenched disorder.  In
samples of La$_{2}$CuO$_{4+\delta}$ prepared by electrochemical
intercalation of oxygen in an aqueous solution, a reversible phase
separation is generally observed.\cite{Wells,Xiong}  In the oxygen-rich
phase, the oxygen interstitials exhibit staging order, indicating
substantial mobility of the oxygens.  Modest reductions in the
superconducting transition temperature, $T_c$, can be achieved by rapid
cooling.\cite{Chou2}  In contrast, crystals intercalated electrochemically
in a molten salt at relatively high temperature can exhibit a large
distribution of lattice parameters at room temperature,\cite{Barilo}
indicative of quenched disorder.  Previous studies\cite{Balag,Pomja} of
such crystals have indicated low oxygen mobility and depressed $T_c$s of
15 to 25~K.

The sample that we study here has, as we will demonstrate, a fine-grained
mixture of oxygen-poor and oxygen-rich domains, with substantial strain,
at 300 K and below.  Neutron diffraction indicates the gradual onset of
antiferromagnetic order below 210~K, while $T_c$ is measured to be 24~K.
These low transition temperatures, together with the absence of
staging order of the interstitials in the oxygen-rich phase, indicate
that quenched disorder has pinned the interstitial distributions within
the first miscibility gap.\cite{Birgeneau}

The first surprise that we observe is that the Raman signature of an
insulating, antiferromagnetically-correlated phase does not appear until
the sample is cooled below $T_r\approx250$~K, even though a significant
degree of oxygen phase separation is already evident.  As we will
discuss, this result suggests that, contrary to expectations, the
miscibility gap is driven largely by lattice, rather than electronic,
energies.  Such behavior is consistent with a study of miscibility gaps
in the related system La$_{2-x}$Sr$_x$NiO$_{4+\delta}$, where the
structural phase diagram is found to depend only on the concentration of
interstitials and not on the net hole concentration.\cite{huck03}

The second surprise is a reduction of the antiferromagnetic Bragg peak
intensity on cooling through $T_c=24$~K.  A somewhat similar reduction in
intensity has been observed in association with an electronic
phase-separation phenomenon in La$_{2-x}$Sr$_{x}$CuO$_4$ with
$x\sim0.015$.\cite{Matsuda} In the latter case, the lost intensity
reappears as diffuse incommensurate scattering; however, we do not see
such a response in our oxygen-doped sample.  Instead, we suggest that the
measurement indicates a reduction in the net volume of antiferromagnetic
phase due to a superconducting proximity effect.  Such a proximity effect
may be necessary for the superconducting phase to propagate through the
finely mixed insulating and metallic phases.

Behavior of this type, associated with quenched structural disorder, has
been discussed recently by Dagotto and collaborators.\cite{Dagotto,Burgy}
Their model calculations show some general features of competition
between two ordered  phases placed on the background of intrinsic
inhomogeneities.  Extended to manganites, this model predicts a colossal
magnetoresistance effect. In the case of underdoped cuprates at low
temperatures with competing superconductivity and antiferromagnetism, the
authors claim that one can expect a ``colossal superconducting proximity
effect.''\cite{Burgy}  Previously, such a phenomenon has been observed in
underdoped YBa$_{2}$Cu$_{3}$O$_{6+\delta}$.\cite{Decca}

The rest of the paper is organized as follows.  A description of
experimental details is given in the next section.  The results are
presented and discussed in Sec.~III.  Our conclusions are summarized in
Sec.~IV.

\section{EXPERIMENTAL DETAILS}

As is now well known, the  way of
preparing La$_{2}$CuO$_{4+\delta}$ plays an important role with
regard to  the physical properties of this material and more
especially its superconducting properties. Electrochemical
oxidation has been shown\cite{Grenier} to be a reliable and
controlled method to induce an oxygen surplus. Using this  process,
oxygen atoms can be inserted into La$_{2}$CuO$_{4}$ at
interstitial positions of the La$_2$O$_2$ layers leading to the
formation of mobile holes of concentration  $p \approx 2\delta$ in the
CuO$_2$ planes. Our crystal was prepared differently from the usual
electrochemical-oxidation method in which the intercalation process occurs
in an aqueous solution at a temperature of 100$^\circ$C and
lower.\cite{Grenier,Chou1,Radael1,Wells,Xiong}  The
oxygenation of our sample was performed electrochemically in a NaOH melt
at $T = 330^\circ$C, with an  electrical current density of 10
mA/cm$^2$ for $\sim47$ minutes.  These are the
same  conditions used for crystal L2 in Ref.~\onlinecite{Zakh}.  The
oxygen excess $\delta$ was estimated to $\sim0.05$.

\begin{figure}[t]
\centerline{\includegraphics[width=7cm]{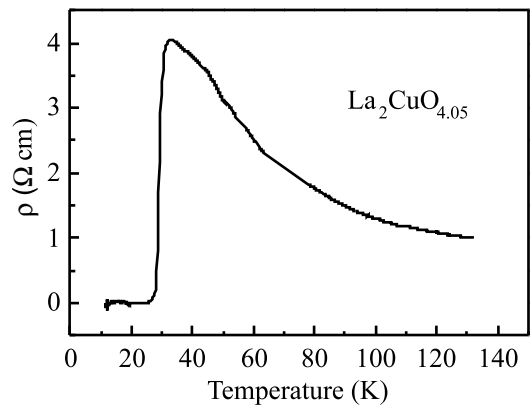}}
\medskip
\caption{Temperature dependence of in-plane resistivity in the
La$_{2}$CuO$_{4.05}$ measured along CuO$_2$ plane.} \label{Resist}
\end{figure}

In-plane resistivity measurements on our sample, Fig.~\ref{Resist}, show
semiconducting behavior.  If this were a single-phase sample, then given
the large magnitude of the resistivity at $T>T_c$, one would not expect it
to go superconducting.\cite{fuku96,semb01} The crystal was also
characterized by SQUID magnetometry. The
superconducting transition temperature was found to be $T_c = 24$~K (see
inset in Fig.~\ref{RamanMain}).

Elastic neutron scattering measurements were performed on the BT-2
triple-axis spectrometer at the NIST Center for Neutron Research. 
The sample was approximately 0.2 cm$^3$ in size.  
After orienting the
crystal on a holder, it was put into an Al can with  He exchange  gas,
and then mounted on the cold  finger  of  a closed-cycle He 
refrigerator. All of the diffraction  measurements were  performed using
an incident neutron energy of 13.7~meV, with either one or two pyrolytic
graphite filters in the incident beam to suppress neutrons at harmonic
wavelengths.  The resolution was varied by changing the horizontal
collimators, as necessary.

Raman scattering measurements were carried out in quasi-backscattering
geometry using the 514.5~nm argon-laser line. A piece of the same
sample used for the neutron measurements was mounted on the holder of a
He-gas-flow cryostat.  The incident laser beam of 10 mW power was focused
onto a spot of 0.1 mm diameter
on the $3\times3$~mm$^2$ sample surface area.
The scattered light was analyzed with
a DILOR XY triple spectrometer combined with a nitrogen-cooled  CCD
detector.

To check the influence of the sample surface condition, the Raman
measurements were performed both on the as-grown surface and on the
mirror-like surface obtained by polishing and chemically etching in
isopropanole and acetone.  As no significant differences were observed,
we report here the measurements performed on the
polished surface.  We note that the penetration depth of the laser light
is about 2000~\AA\ in cuprate superconductors.\cite{thom89,ShermanRev}

The  crystal  structure of stoichiometric La$_{2}$CuO$_{4}$ is
orthorhombic ($D_{2h}^{18}$) at room temperature and tetragonal
($D_{4h}^{17}$) above about 515K. The  orthorhombic distortion is
small and this allows us  to assume tetragonal  symmetry  as is
usually done in Raman studies of La-cuprates.\cite{Coop1,Coop2}
The $x$- and $y$-axes were taken to be along the CuO bonds, with $z$
perpendicular to the CuO$_2$ plane. Within the tetragonal point
group $D_{4h}$, the $zz$ geometry couples to excitations  of
$A_{1g}$ symmetry, $xy$ to $B_{2g}$ symmetry, and $xx$ to the
combination of $A_{1g}$  and $B_{1g}$ symmetries. If $x'$ and $y'$
denote axes rotated by 45$^\circ$ from $x$ and $y$, then $x'y'$
geometry allows coupling to excitations with $B_{1g}$ symmetry and
$x'x'$ geometry to a combination of $A_{1g}$ and $B_{2g}$
components. The spectra reported here were measured in the $xx$, $xy$,
$x'x'$,  and $x'y'$ scattering configurations.

\section{RESULTS AND DISCUSSIONS}

\subsection{Structural data}

For single crystals prepared electrochemically in an aqueous solution of
NaOH, an ordering of the oxygen interstitials in oxygen-rich  phases  has
been observed.\cite{Wells,Xiong}  The  interstitials separate
into domains such that, in a given domain,  interstitials are
present only in every $n^{\rm th}$ layer.  This ``staged''  structure  is
easily identified by the presence of specific superlattice peaks.
We expected  that it should be possible to identify the oxygen
content of the intercalated phase by characterizing the superlattice
peaks.  Surprisingly, neutron  diffraction  measurements  on  the
present crystal revealed  an  absence of such  superlattice peaks
at all temperatures.  On the other hand, measurements of
fundamental Bragg peaks  indicate the presence of two phases, with
lattice parameters consistent with oxygen-rich and oxygen-poor
phases.

\begin{figure}[t]
\centerline{\includegraphics[width=8cm]{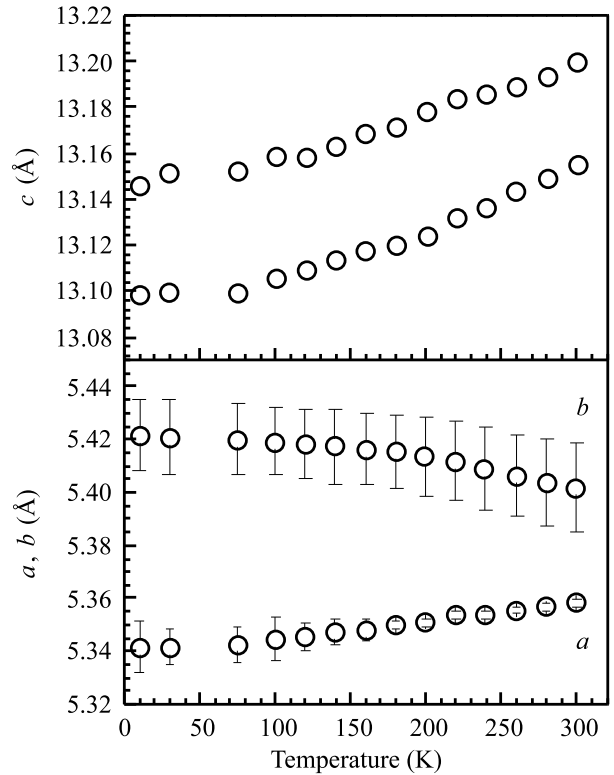}}
\medskip
\caption{Temperature dependence of the $c$ (top) and of the $a$ and
$b$ (bottom) lattice parameters.
Error bars in (b) indicate peak widths (corrected for resolution).}
\label{LatticeCon}
\end{figure}

The presence of two phases is most clearly indicated by $(00l)$
reflections, and we have focused on the (006).  Although two distinct
peaks cannot be resolved, the observed peak is quite broad and
asymmetric.  Fitting with two gaussian peaks yields the $c$ lattice
parameters shown in Fig.~\ref{LatticeCon}(a).  These results are
consistent with previous studies of phase-separated
samples.\cite{Radael2,Wells} The phase with the larger $c$ lattice
parameter should be the oxygen-rich phase. The difference between lattice
parameters changes little with temperature, suggesting that the oxygen
phase separation temperature, $T_{ps}$, is well above 300~K.

\begin{figure}[t]
\centerline{\includegraphics[width=7cm]{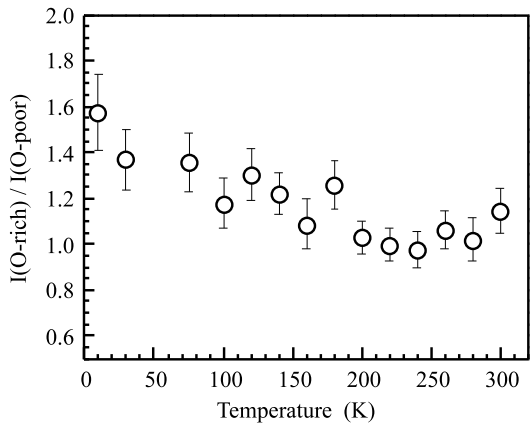}}
\medskip
\caption{Ratio of integrated intensities of the two (006) peaks vs.\
temperature.} \label{RatioPhases}
\end{figure}

The relative intensities of the two fitted (006) peaks are shown as a
function of temperature in Fig.~\ref{RatioPhases}.  Assuming similar
structure factors, there are comparable amounts  of  the  two phases,
with slightly more of the oxygen-rich phase. The temperature dependence
of the relative intensities  might indicate  changes  in  the volume
fractions; however,  it  is  also possible  that  the  changes are  due
to relief  from  extinction associated with strain. For the
oxygen-rich phase, the widths of the (002), (006), and (008) Bragg
peaks increase with momentum transfer, indicating a substantial
amount of strain. This strain is expected to have an impact  on the  phase
separation and oxygen ordering  dynamics;  it indicates a significant
amount of quenced disorder.
If we assume that the fitted peak widths obtained at (002) are due purely
to particle-size broadening (ignoring resolution and strain), then we get
a lower limit of 600~\AA\ for the diameters of the oxygen-poor and
oxygen-rich domains.

A scan through (200)/(020) shows two peaks.  (These reflections are seen
along the same direction because of twinning.)  One should expect to see
three  peaks  due  to the difference in  orthorhombicity found
previously  for the oxygen rich and poor phases.\cite{Radael2}  The
observed peaks  are  slightly asymmetric at low temperature, but a fit
with more than two peaks is not stable over a broad temperature  range.
The lattice parameters obtained from the 2-peaks fits are shown in
Fig.~\ref{LatticeCon}(b).  The error bars reflect the  distribution
of  lattice parameters determined from the peak widths after
correcting for the calculated resolution. The width of the $b$
distribution is comparable to the splitting in $b$ for the two
phases found by Radaelli \emph{et al}.\cite{Radael2}  The
temperature dependence of structural superlattice peaks at (012)
appears  to be consistent with a very gradual  decrease  in
orthorhombic splitting with increasing temperature.

\subsection{Onset of insulating state and features of magnetic
order}

\begin{figure}[t]
\centerline{\includegraphics[width=7cm]{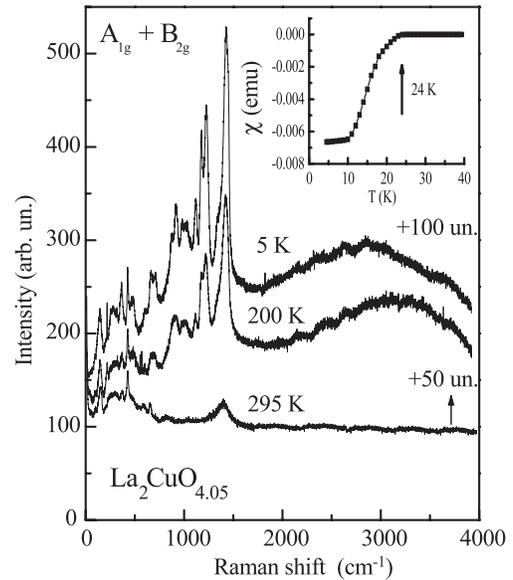}}
\medskip
\caption{Raman spectra of La$_{2}$CuO$_{4.05}$ at 5, 200, and 295~K in
$x'x'$ polarization configuration.  The inset shows the magnetic
susceptibility $\chi$ of the sample.}
\label{RamanMain}
\end{figure}

\begin{figure}[b]
\centerline{\includegraphics[width=7cm]{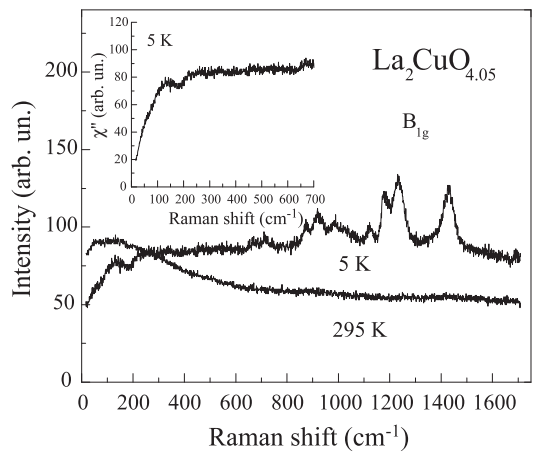}}
\medskip
\caption{Raman spectra of La$_{2}$CuO$_{4.05}$ at 5 and 295~K in $x'y'$
polarization configuration.  The inset shows the Raman response function
$\chi''(\omega)$ obtained by dividing the original spectra
by the Bose-Einstein thermal factor $[1 - \exp(-h\omega/k_{B}T)]^{-1}$.
The downturn at low frequency is a signature of the superconducting
state.\cite{Chen3}
 }
\label{RamanB1g}
\end{figure}

Raman spectra of La$_{2}$CuO$_{4.05}$ at different temperatures
in  the $x'x'$ and $x'y'$ polarization configurations are shown in
Figs.~\ref{RamanMain} and \ref{RamanB1g}, respectively.  For both
configurations, one can see very strong two-phonon scattering at low
temperature, but very little at 295~K.  (Note that there are no single
phonon excitations above $\sim700$~cm$^{-1}$.)  In addition, one can see
strong two-magnon scattering\cite{lyon88} of $A_{1g}$ symmetry at
$\sim3000$~cm$^{-1}$ in the $x'x'$ configuration when the two-phonon
scattering is strong.

\begin{figure}[t]
\centerline{\includegraphics[width=6cm]{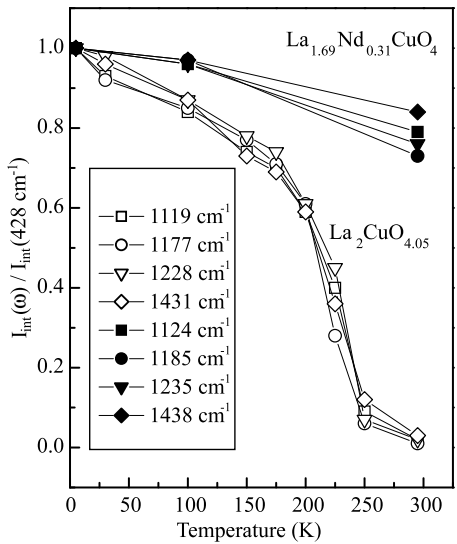}}
\medskip
\caption{Integrated scattering intensities of the two-phonon lines
normalized to the intensity of the 428 cm$^{-1}$ phonon peak as a function
of temperature in La$_{2}$CuO$_{4.05}$ and
La$_{1.69}$Nd$_{0.31}$CuO$_{4}$. } \label{RamanPhSep}
\end{figure}

In Raman studies of La$_{2-x}$Sr$_x$CuO$_4$, Sugai and
coworkers\cite{Sug89,Sug90} have shown that strong two-magnon and
two-phonon scattering peaks appear together in the
antiferromagnetic-insulator phase at
$x\approx0$, but both are dramatically reduced in doped phases with
$x\gtrsim0.06$.  Thus, it appears that the two-phonon signal, along with
the two-magnon scattering, is a useful indicator of the insulating
phase.  To test the temperature dependence of this signature, we
performed measurements on the insulator
La$_{1.69}$Nd$_{0.31}$CuO$_{4}$.\cite{Wagener}  The temperature-dependent
intensities of several two-phonon peaks, relative to the intensity of the
single-phonon peak at 428~cm$^{-1}$, are plotted in
Fig.~\ref{RamanPhSep}.  We see that there is only a modest decrease in
the intensity ratio between 5~K and 295~K.

The two-phonon intensity ratios for the La$_2$CuO$_{4.05}$ sample are
also plotted in Fig.~\ref{RamanPhSep}.  Compared to the insulator, the
relative two-phonon signal decreases more rapidly with increasing
temperature, dropping rapidly toward zero near 250~K.  We interpret these
results as indicating the absence of an insulating phase for
$T>T_r\approx250$~K, and the rapid appearance of the insulating phase for
$T<T_r$.

\begin{figure}[t]
\centerline{\includegraphics[width=7cm]{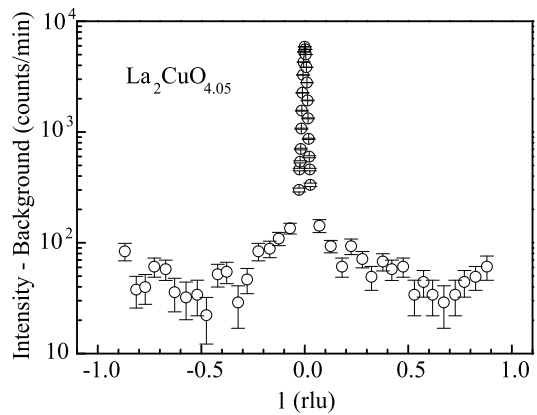}}
\medskip
\caption{Scan along ${\bf Q} = (1,0,l)$ through the magnetic Bragg
peak at $l=0$, measured at $T = 9$ K.
Background signal has been subtracted, and intensity is on a logarithmic
scale.} \label{ENSmagnSign}
\end{figure}

\begin{figure}[b]
\centerline{\includegraphics[width=8cm]{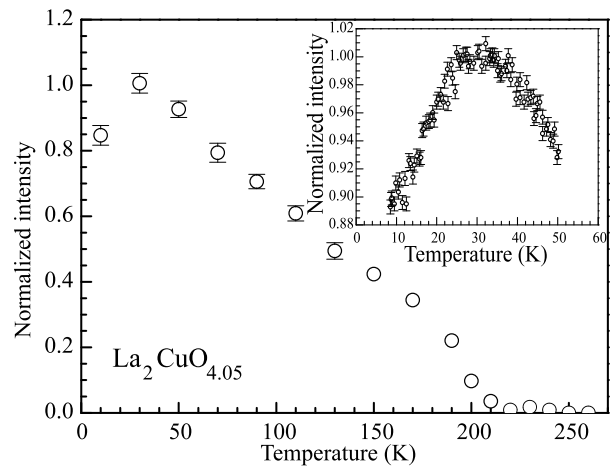}}
\medskip
\caption{Temperature dependence of the (100) magnetic Bragg peak
intensity. The inset shows the intensity of the same peak at low
temperatures.}
\label{ENSmagnSuppr}
\end{figure}

To detect the onset of antiferromagnetic order, we rely on neutron
scattering measurements.  Figure~\ref{ENSmagnSign} shows a scan along
${\bf Q}=(1,0,l)$ through the magnetic Bragg peak at $l=0$, measured at $T
= 9$~K.  Besides the Bragg peak, there is also broad diffuse scattering
along  the antiferromagnetic rod.  The temperature dependence of the (100)
magnetic peak is shown in Fig.~\ref{ENSmagnSuppr}.  The N\'eel temperature
appears to be approximately 210~K, which is significantly lower than
T$_r$.  This magnetic ordering temperature is also lower than the more
typical value of $\sim250$~K found for the hole-poor phase in two-phase
samples,\cite{JMT_rev} thus indicating that the quenched disorder causes
some frustration of the phase separation.

\subsection{Analysis of the phase separation transitions}

The neutron diffraction measurements indicate the presence
of two phases near the two  edges of the  first  miscibility
gap\cite{Radael2} all the way up  to 300~K.  In previous studies, the
oxygen-poor phase has been identified as an antiferromagnetic insulator,
based on detection of antiferromagnetic order at lower temperatures;
however, in our case the Raman spectra indicate that an antiferromagnetic
insulating phase does not appear until the temperature drops below
$T_r$.  How can we reconcile these observations?

Due to long-range Coulomb interactions, it is clear that some degree of
electronic phase separation must occur concomitantly with the
development of two phases with different interstitial concentrations.  At
the same time, the antiferromagnetic phase of lightly-doped La$_2$CuO$_4$
is quite sensitive to hole content, with $T_N$ dropping rapidly to zero
at $p=0.02$ and the transport properties varying rapidly with $p$ in this
regime.\cite{Kastner,Ando}  Hence, the oxygen-poor phase at $T>T_r$
may correspond to $p\gtrsim0.02$.  On cooling through $T_r$, the
electronic energies favoring expulsion of holes from the
antiferromagnetic phase\cite{Emery} become important and drive a slight
increase in the degree of phase separation, resulting in the onset of
detectability of signatures of the antiferromagnetic phase.  We presume
that the slight adjustment of interstitial segregation at this point was
masked in the diffraction experiments by the large strain effects.

The idea that the initial, high-temperature phase separation is driven
largely by lattice energies rather than electronic energies is consistent
with a recent study\cite{huck03} of room-temperature structural phase
diagrams in the related system La$_{2-x}$Sr$_x$NiO$_{4+\delta}$.  With
$x=0$, pure La$_2$NiO$_{4+\delta}$ is known to have a complicated phase
diagram with multiple miscibility gaps.\cite{JMT_rev}  If these phase
separations were driven dominantly by electronic energies, then on
co-doping with Sr, one would expect the phase boundaries to scale with
the net hole content,
$p=x+2\delta$.  To the contrary, diffraction measurements on numerous
samples clearly show that the phase boundaries depend almost exclusively
on $\delta$ (at least for $x<0.08$).  Thus, it appears that the lattice
strain energies associated with interstitial oxygens must play an
important role in the phase separation phenomena.

\subsection{Competition between superconductivity and antiferromagnetism}

The variation  of the (100) magnetic peak intensity at low temperature is
shown in the inset of Fig.~\ref{ENSmagnSuppr}.  On cooling, the
intensity appears to hit a plateau at about 35~K, and then below
$T_c$ it decreases.  This effect looks similar to that
observed\cite{Matsuda} in antiferromagnetic La$_{2-x}$Sr$_x$CuO$_4$  with
$0.01 < x < 0.02$, and in oxygenated La$_2$CuO$_4$ with extremely
low N\'eel temperature ($T_N\sim 90$~K).\cite{suppression2,suppression1}
However, in contrast to those observations, a corresponding
increase in the diffuse magnetic scattering was not observed in our case.

The drop in the (100) magnetic peak intensity at low temperature is quite
remarkable.  We propose that  the observed effect may involve a change
in the relative volume fraction of metallic and insulating phases for
temperatures below the superconducting phase transition. At such
low temperatures the mobility  of excess oxygen is negligible, so
the changes should be largely electronic in nature.  It appears that a
form of superconducting proximity effect may be involved.  The model
calculations of Burgy \emph{et al.}\cite{Burgy} indicate that such a
phenomenon may occur when a system with quenched disorder is below a
normal-state percolative threshold.  We note that the in-plane resistivity
of our crystal (Fig.~\ref{Resist}) looks qualitatively quite similar to
calculated curves in Fig.~4(b) of Ref.~\onlinecite{Burgy}.

\subsection{Features of the superconducting state}

Since  the discovery of high-temperature superconductivity,  many
experimental  techniques,  such as quasiparticle  tunneling,
angle-resolved photoemission,  microwave  absorption,   IR and
Raman spectroscopy,  etc., have been used with the aim  to clear
up  the nature   of   the   superconducting  gap  (SCG). Raman
scattering established  itself as a powerful method to study this
problem.  At room  temperature  all  cuprates exhibit a  broad and
rather  flat background that is usually attributed to electronic
excitations.  As the  temperature  decreases below T$_c$, the
frequency distribution  of the electronic background changes
reflecting the opening of the SCG. This  renormalization consists
of the formation  of  a  broad  peak associated with the
pair-breaking process at the energy 2$\Delta$  together with the
decrease of the scattering intensity at energies lower than
2$\Delta$. Unlike conventional superconductors with an isotropic
gap, where no scattering is expected for frequencies $\omega \leq
2\Delta$, the low frequency depletion  in  high-temperature
superconductors (HTSC's)  is not complete. Moreover, the peak
frequency positions and renormalization of  the scattering
intensity below $T_c$ is different for $A_{1g}$, $B_{1g}$, and
$B_{2g}$  symmetry components. An appropriate choice of the
polarizations of incident and scattered light in the RS experiment
can reveal the SCG anisotropy and the symmetry of the order
parameter.

Most  of  the  RS  experiments imply that the SCG  in  HTSC's  is
strongly anisotropic.\cite{Coop1,Coop2,Hackl88,Hackl89,Stauff,%
Dever94,Chen1,Chen2,Chen3,Chen4,Gasp,Chen5,Misoch}.
Such a behavior was examined by different authors in the framework of a
$d$,\cite{Scalap,Month} $s + id$,\cite{Kotliar} or anisotropic
$s$-pairing.\cite{Chakr}  On the other hand, some experiments performed
mainly on single CuO$_2$-layered compounds\cite{Mason,JChen,Anlage,Stadl}
suggest the presence of an isotropic $s$-wave type gap. Up to now there
is no definite consensus  how  to interpret these conflicting results.
Furthermore, it is not clear whether the SCG anisotropy depends on
doping  level.  Raman scattering experiments on
Tl$_2$Ba$_2$CuO$_{6+\delta}$ and Bi$_2$Sr$_2$CaCu$_2$O$_{8+\delta}$
showed that overdoped samples exhibit a symmetry independent gap in
contrast to the anisotropic SCG in the near optimum doping
case.\cite{Kendz}  A SCG anisotropy was observed in
YBa$_2$Cu$_3$O$_{7-\delta}$ with different doping
levels.\cite{Chen6,Kall}  The anisotropy of the SCG and variation of the
maximum in the electronic RS with scattering selection rules have been
discussed in detail in a recent paper of M. Opel \emph{et al}.\cite{Opel}
It was pointed out that $d$-wave symmetry is widely accepted but
complications arise both in the underdoped and overdoped ranges of the
phase diagram. It is thus of interest to check the redistribution of the
electronic continuum below $T_c$ in the metallic phase of the sample that
undergoes phase separation.

\begin{figure}[t]
\centerline{\includegraphics[width=7cm]{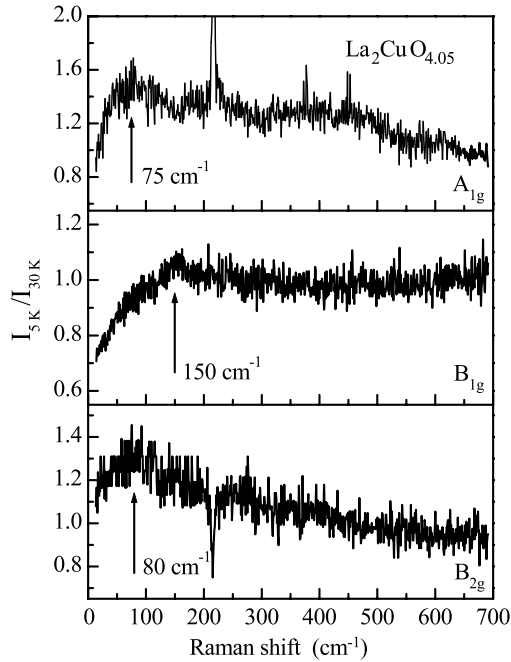}}
\medskip
\caption{Electronic Raman scattering of La$_{2}$CuO$_{4.05}$. Shown are
normalized spectra, $I(T=5 K)/I(T=30 K)$.} \label{RamanSPgap}
\end{figure}

In our experiments on La$_{2}$CuO$_{4.05}$ a frequency
redistribution of the  electronic  continua is observed in all
scattering geometries investigated when the temperature decreases
below $T_c=24$~K. In  order to  evaluate  the  $A_{1g}$ scattering
component, we subtracted  the  $xy$ spectra from the $xx$ spectra. To
emphasize the redistribution of  the scattering intensity in the
superconducting state compared  to  the normal state,  the spectra
at 5 K were divided by the  spectra  just above $T_c$. As one can
see from Fig.~\ref{RamanSPgap}, the low-frequency behavior of the
electronic Raman scattering in La$_{2}$CuO$_{4.05}$ exhibits  a
strong anisotropy  with respect to the symmetry components.  The
pair-breaking peaks are located at 75, 150, and 80 cm$^{-1}$ for
the $A_{1g}$, $B_{1g}$  and  $B_{2g}$ geometries,
respectively.  As was described in  detail
earlier,\cite{Dever94,Gasp,Dever95} the symmetry of the order parameter
can  be inferred from  the specific spectral features for  each
scattering component.  To our knowledge there are  no results published
on electronic Raman scattering in oxygen doped La$_{2}$CuO$_{4+\delta}$.
Therefore, we want  to compare  our data with experiments on  Sr-doped
La$_2$CuO$_4$.\cite{Chen3,Chen5,Misoch}

Electronic RS studied\cite{Chen3,Chen5} in optimally doped
La$_{2-x}$Sr$_x$CuO$_4$ supports the existence of an anisotropic SCG.
Besides that, it was observed that the intensities in the low
frequency portions of the $B_{1g}$  and $B_{2g}$  continua
increase as $\omega^3$  and $\omega$, respectively,  in
quantitative agreement with the $d$-wave hypothesis.  Electronic RS
in La$_{2-x}$Sr$_x$CuO$_4$ with various levels of doping was
studied  in Ref.~\onlinecite{Misoch}. Theere, the anisotropy concerning
the symmetry dependent energies of the maximum was confirmed; however,
the $\omega^3$ low-frequency behavior was absent at all doping
levels.  This behavior can be attributed to resonance effects in Raman
scattering.\cite{Sherman}

We do not want to advocate here whether a  $d$-wave or
anisotropic $s$-wave pairing is appropriate to describe
La$_{2}$CuO$_{4.05}$, but it should be noted that we also do not
observe a $\omega^3$ law in the $B_{1g}$ scattering component (see
Fig.~\ref{RamanSPgap} and inset in Fig.~\ref{RamanB1g} ). The data
shown here are very consistent with the experimental results of
Ref.~\onlinecite{Misoch}; however, in our study the situation is
complicated due to phase separation. As discussed above and shown
in Fig.~\ref{RamanMain}, for temperatures below $T_c$ we detect
both the Raman signal characteristic for  the insulating as well
as the superconducting phases. Moreover, we expect the oxygen
concentration of the oxygen-rich regions to vary significantly,
consistent with the strain detected by neutron diffraction.

\section{CONCLUSION}

In  summary,  we  have  carried out  Raman  and  elastic  neutron
scattering  experiments  on an excess-oxygen-doped
La$_{2}$CuO$_{4}$ single crystal.  The neutron diffraction
measurements indicate that  there are both oxygen-rich and
oxygen-poor phases present in the sample and that the oxygen phase
separation temperature is well above 300 K. The absence of
detectable staging behavior in oxygen-rich regions allows us to
conclude that a significant amount of quenched disorder must be present in
the sample. In the Raman scattering experiments,
the La$_{2}$CuO$_{4.05}$ sample displays at room temperature a response
similar to cuprates with substantial doping. Below the temperature
$T_r\approx250$~K, the RS spectra change drastically: features
characteristic for the insulating state appear in the spectra.  We have
proposed that electronic energies cause an additional redistribution of
interstitial oxygen at $T_r$, making the oxygen-poor phase more
insulating and resulting in the appearance of the two-phonon and
two-magnon scattering. The N\'eel temperature $T_N \approx 210$~K was
estimated from the neutron scattering measurements. A decrease of the
(100) magnetic Bragg peak intensity for  $T<T_c$ is attributed to
subsequent electron-density redistribution connected with a
superconducting proximity effect.  Below the superconducting transition
temperature, two-magnon scattering, strong two-phonon scattering, and a
change of the electronic continua (that occurs as a result of the
superconducting gap opening) are observed in the  RS spectra,
simultaneously.  Based  on  the polarization dependence of the
electronic  continua, we conclude that the SCG in
La$_{2}$CuO$_{4.05}$  is anisotropic with an anisotropy ratio
$\Delta(B_{1g})/\Delta(B_{2g})\approx 1.9$ for the CuO$_2$
plane. The peculiarities of  the electronic  RS,  i.e., the low-
frequency power   law dependence  of  the  different scattering
components at low frequencies and their different peak positions,
are found to be very close to those observed earlier for
La$_{2-x}$Sr$_x$CuO$_4$.\cite{Misoch}


\section*{Acknowledgments }

We thank E. Ya.\ Sherman for helpful discussions.  This  work was 
supported  in  part  by  NATO Collaborative Linkage Grant PST.CLG.977766
and INTAS Grant 96-0410. JMT  is supported at Brookhaven by the U.S.
Department of  Energy's Office of  Science  under  Contract  No.
DE-AC02-98CH10886.    We acknowledge  the  support  of the National
Institute of Standards and Technology, U.S. Department of Commerce, in
providing the neutron facilities used in this work.

\end{document}